\documentclass[runningheads]{llncs}
\usepackage[T1]{fontenc}
\usepackage{graphicx}
\usepackage{amsmath}
\usepackage{amssymb}
\usepackage{booktabs}
\usepackage{array}
\usepackage{url}
\usepackage{subcaption} 
\usepackage{hyperref}
\usepackage{xurl}
\usepackage{color}

\begin{document}
%
\title{Approximating SPR Distance Between Phylogenetic Trees with Graph Neural Networks}
\titlerunning{Learning Phylogenetic Tree Similarity with GNNs}
\author{Renata Martins Castanheira\inst{1}\orcidID{0009-0005-6690-6707} \and
Miguel Bugalho\inst{1,2}\orcidID{0000-0002-4630-9615} \and
Cátia Vaz\inst{1,3}\orcidID{0000-0001-6074-3074}}
\authorrunning{R. M. Castanheira et al.}
\institute{ISEL -- Instituto Superior de Engenharia de Lisboa,
Instituto Politécnico de Lisboa, Portugal \and
ALGORITMI Research Center/CCPM, Portugal \and
INESC-ID Lisboa, Portugal \\
\email{51830@alunos.isel.ipl.pt}, \email{miguel.bugalho@isel.pt}, \email{cvaz@cc.isel.ipl.pt}}
\maketitle
\pagenumbering{gobble}   
\begin{abstract}
Comparing phylogenetic tree topologies is essential for understanding
epidemic dynamics, yet biologically meaningful distances such as the
Subtree Prune and Regraft (SPR) distance are NP-hard to compute and
intractable on large datasets. We investigate whether a Graph Neural
Network (GNN) can approximate SPR distances in near-constant time per
comparison after training. Our contributions are fourfold. First, we
build and publicly release a dataset of 864 phylogenetic trees inferred
with UPGMA and Neighbor-Joining over four bacterial species, spanning up to 9{,}500 isolates,
together with 388 labelled tree pairs. Second, we establish a
reproducible pre-processing pipeline including midpoint re-rooting, which
reduces tree depth and supplies the rooting required for exact distance
computation and for the model's root-based features. Third, we validate
the supervision target: on small trees, where exact SPR is tractable, the
unrooted \texttt{phangorn::SPR.dist} heuristic correlates almost perfectly
with the exact rooted distance computed by \texttt{rspr}
(Pearson $0.98$--$0.99$), making it an excellent monotonic surrogate.
Lastly, we train a Siamese Graph Isomorphism Network (GIN) regressor.
In-distribution, {em i.e.}, held-out trees from the same species and size range as training, it explains roughly 87--90\% of the variance
($R^2 \approx 0.87$ on a held-out split; $0.90 \pm 0.19$ under stratified
cross-validation), with about four times lower error than a mean-predictor
baseline, and shows partial transfer to unseen species
($R^2 \approx 0.37$). 
Its main limitation is extrapolation to trees larger than those seen in
training, where accuracy collapses. The released dataset and the
validated heuristic versus exact relationship provide a reproducible basis for
scaling learned SPR approximation.
\end{abstract}
\section{Introduction}
\label{sec:intro}
Determining the origin, evolutionary dynamics and transmission patterns
of bacterial and viral epidemics is among the central challenges of
computational biology. Phylogenetic inference reconstructs these
evolutionary relationships. Its outcome, the phylogenetic
tree~\cite{kapli2020}, is an indispensable data structure for tasks such
as tracking pathogen strains through an outbreak.

Different inference methods, such as Maximum Likelihood~\cite{felsenstein1981},
Maximum Parsimony~\cite{fitch1971}, Neighbor-Joining
(NJ)~\cite{saitou1987}, UPGMA~\cite{sokal1958}, or
goeBURST~\cite{francisco2009goeburst}, frequently produce distinct
topologies for the same isolates~\cite{vaz2021distance}, so evaluating and
reconciling them requires comparing trees at scale. The available distance
measures trade speed against biological meaning (Section~\ref{sec:related}):
cheap partition-based measures such as Robinson--Foulds~\cite{robinson1981}
saturate quickly and overreact to a single displaced taxon, whereas the
Subtree Prune and Regraft (SPR) distance~\cite{hickey2008} directly counts
the prune-and-regraft operations between two topologies and so models
reticulate events such as recombination and horizontal gene transfer.
That biological fidelity is costly since computing the rooted SPR distance is
NP-hard~\cite{bordewich2005}, which rules out exact computation for trees
with thousands of leaves.

This motivates an approximation that is both accurate and fast. Since the exact rooted distance is intractable at scale, it cannot serve as a training signal for large trees; we therefore adopt a polynomial-time heuristic, the unrooted phangorn::SPR.dist, as the supervision target, after first verifying on small trees—where exact computation is feasible—that it tracks the exact rooted SPR distance almost perfectly (Pearson 0.98–0.99). We then ask whether a GNN~\cite{wu2021}  can learn to reproduce this surrogate directly from tree structure, so that each comparison reduces to a single forward pass at inference time. Our contributions are:
\vspace{-0.2cm}
\begin{enumerate}
\item A reproducible pre-processing pipeline that slices genomic
profiles into size-stratified sub-datasets, diversifies them through
controlled shuffling, infers UPGMA and NJ trees, applies midpoint
re-rooting~\cite{mai2017}, and extracts SPR supervision labels.
\item A publicly released dataset~\cite{castanheira2026} of 864 trees and
388 labelled tree pairs across four bacterial species, with isolates up
to 9{,}500 sequences.
\item A validation of the supervision target against the exact rooted SPR
distance computed with \texttt{rspr}~\cite{whidden2014}, clarifying the
relationship between the exact metric and the
\texttt{phangorn}~\cite{schliep2011} heuristic.
\item A Siamese GIN~\cite{xu2019} regressor with training and inference
pipelines\footnote{publicly available at
\url{https://github.com/RenataCastanheira/LearningPhylogeneticTreeSimilarityWithGraphNeuralNetworks}}, together with an assessment of where it works (in-distribution), where transfer is partial (across species), and where it fails (size extrapolation).
\end{enumerate}
\vspace{-0.6cm}
\section{Related Work}
\label{sec:related}

Molecular epidemiology and comparative genomics increasingly reconstruct
phylogenies from typing and whole-genome data for collections reaching
thousands of isolates, where the resulting trees drive outbreak
investigation and the study of recombination and horizontal gene
transfer~\cite{kapli2020}. Because standard inference methods, such as
NJ~\cite{saitou1987}, UPGMA~\cite{sokal1958},
goeBURST~\cite{francisco2009goeburst}, Maximum Likelihood~\cite{felsenstein1981}
and Maximum Parsimony~\cite{fitch1971}, routinely return different
topologies for the same isolates~\cite{vaz2021distance}, comparing trees is
a recurring and increasingly large-scale step in any such analysis. The
metric chosen for that comparison is in practice a tooling decision with
biological consequences. The Robinson--Foulds distance~\cite{robinson1981}
is the field's workhorse because it is linear-time and widely implemented,
but it saturates on divergent isolates and overreacts to a single misplaced
taxon, limiting its use for fine-grained outbreak resolution;
information-theoretic variants soften this~\cite{smith2020}.
Quartet~\cite{estabrook1985} and path-difference~\cite{steel1993} measures
recover more topological detail but report a value that does not translate
cleanly into evolutionary events. Rearrangement distances are the
biologically literate alternative: the Subtree Prune and Regraft (SPR)
distance~\cite{hickey2008}, like the related Tree Bisection and
Reconnection~\cite{allen2001}, counts the edit operations between two trees
and so directly models reticulate processes such as recombination and
horizontal gene transfer. The cost is computational, since rooted SPR is
NP-hard~\cite{bordewich2005}, and this shapes the available software:
\texttt{rspr}~\cite{whidden2014} delivers exact rooted distances through maximum agreement forests but only for
modest trees, whereas \texttt{phangorn}~\cite{schliep2011} scales by
estimating the unrooted distance heuristically, with no approximation
guarantee, based on the algorithm of de Oliveira Martins et
al.~\cite{deoliveira2008}. We address this lack of a guarantee by validating
the heuristic directly against the exact rooted distance
(Section~\ref{sec:eval}) before adopting it as our supervision target. A
related quantity, the lower bound on the unrooted distance, is
similarly tractable via agreement forests~\cite{whidden2019unrooted}. What
no existing tool offers is amortized, near-constant-time comparison: exact
methods do not scale, and although the heuristic scales, it must be
recomputed for every pair.

This is the gap a learned estimator can fill, and graph neural
networks~\cite{wu2021} are a natural instrument because a phylogeny is a
graph. Their behaviour on this data hinges on the aggregation function: the
degree-normalised mean of Graph Convolutional Networks~\cite{kipf2017} and
the learned attention of Graph Attention Networks~\cite{velickovic2018} tend
to smooth away structure on the near-3-regular topologies of binary trees,
whereas the injective sum of Graph Isomorphism Networks~\cite{xu2019}
preserves exactly the differences the SPR distance depends on, matching the
discriminative power of the Weisfeiler--Lehman test.

For pairwise comparison, two paradigms dominate. Cross-graph matching
networks such as Graph Matching Networks~\cite{li2019} inject node-to-node
attention for fine-grained accuracy but scale poorly with graph size. Graph
embedding approaches, exemplified by SimGNN~\cite{bai2019}, encode each graph
independently and predict from the pair of embeddings; these Siamese
designs~\cite{bromley1993signature} are cheaper and scale better. Given our
target of trees with thousands of leaves, we adopt the embedding paradigm
with a Siamese GIN encoder. To our knowledge, approximating the NP-hard SPR
distance on rooted phylogenies with this approach has not previously been
addressed.
\vspace{-0.05cm}
\section{Dataset and Ground-Truth Construction}
\label{sec:dataset}
\begin{figure} 
\centering
\includegraphics[width=\textwidth]{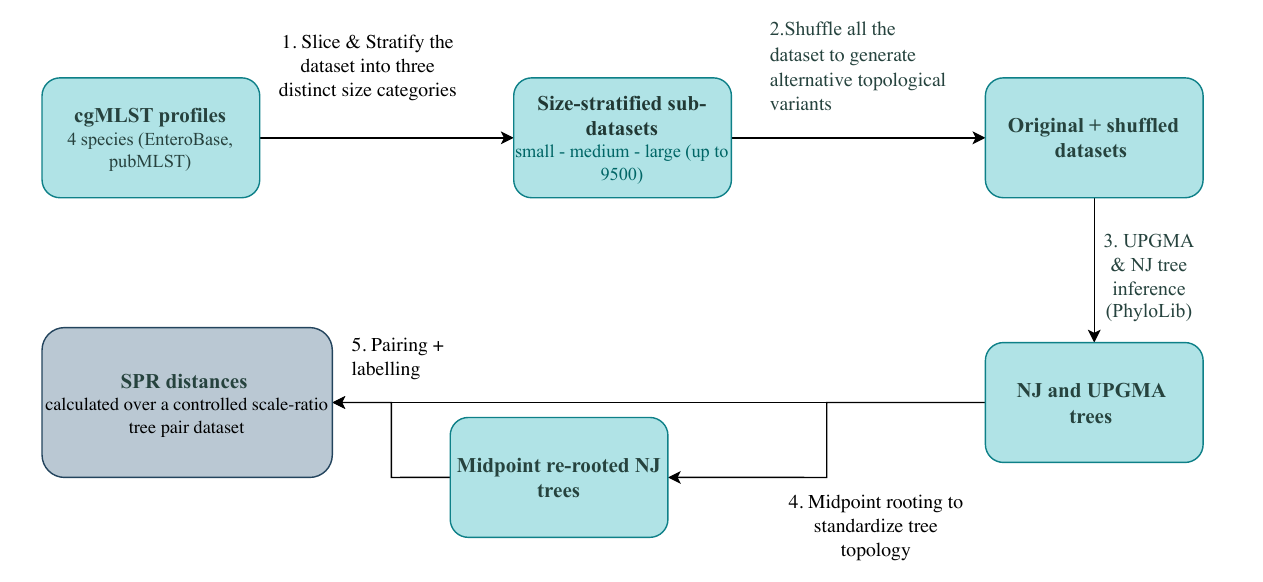}
\caption{Pre-processing pipeline: dataset slicing, controlled shuffling,
UPGMA/NJ inference, midpoint re-rooting, and SPR labelling.}
\label{fig:pipeline}
\end{figure}

As depicted in Figure~\ref{fig:pipeline}, the dataset is organised as a five-stage pipeline.
\subsubsection{Profiles and stratification.}
We sample core-genome MLST (cgMLST)~\cite{maiden2013} profiles for four
bacterial species: \textit{Clostridium}, \textit{Salmonella} and
\textit{Vibrio} from EnteroBase~\cite{dyer2024enterobase}, and
\textit{Streptococcus pneumoniae} from PubMLST~\cite{jolley2018pubmlst}. A custom R script
slices the original profiles sequentially, always from the first row
onward, into three size tiers: \emph{small} (90--1{,}000 isolates, 25\%),
\emph{medium} (2{,}000--6{,}000, 25\%) and \emph{large}
(7{,}000--9{,}500, 50\%); see Fig.~\ref{fig:datasetSizeSplit}. Slicing
contiguously preserves the implicit temporal and geographic structure of
the profiles and ensures smaller sub-datasets are nested within larger
ones.
\subsubsection{Controlled shuffling.}
Distance-based methods such as NJ and UPGMA are sensitive to input order
when breaking ties in the distance matrix. We exploit this with a seeded
permutation script that reorders profile rows, forcing the algorithms to
resolve ties differently and produce topologies with minor, controlled
structural variations.
\subsubsection{Inference and re-rooting.}
From each (original and shuffled) sub-dataset we compute a Hamming
distance matrix and infer UPGMA and NJ trees in Newick format with
PhyloLib~\cite{phylolib2026}. NJ trees are unrooted, with a basal
trifurcation, so we apply midpoint rooting~\cite{mai2017} using
Biopython's \texttt{Bio.Phylo}~\cite{cock2009}. This yields the rooted
binary trees required by the exact rooted-SPR algorithm and defines the
model's root-based node features. Rooting also balances the topology: on
a 469-isolate \textit{Salmonella} tree it reduced maximum topological
depth from 52 to 33 and average depth from $\approx 31.2$ to
$\approx 16.0$, which lowers the cost of the traversal-based heuristic
label computation.
\subsubsection{Pairing and labelling.}
\begin{figure}
    \centering
    \begin{subfigure}[t]{0.48\textwidth}
        \centering
        \includegraphics[width=\textwidth]{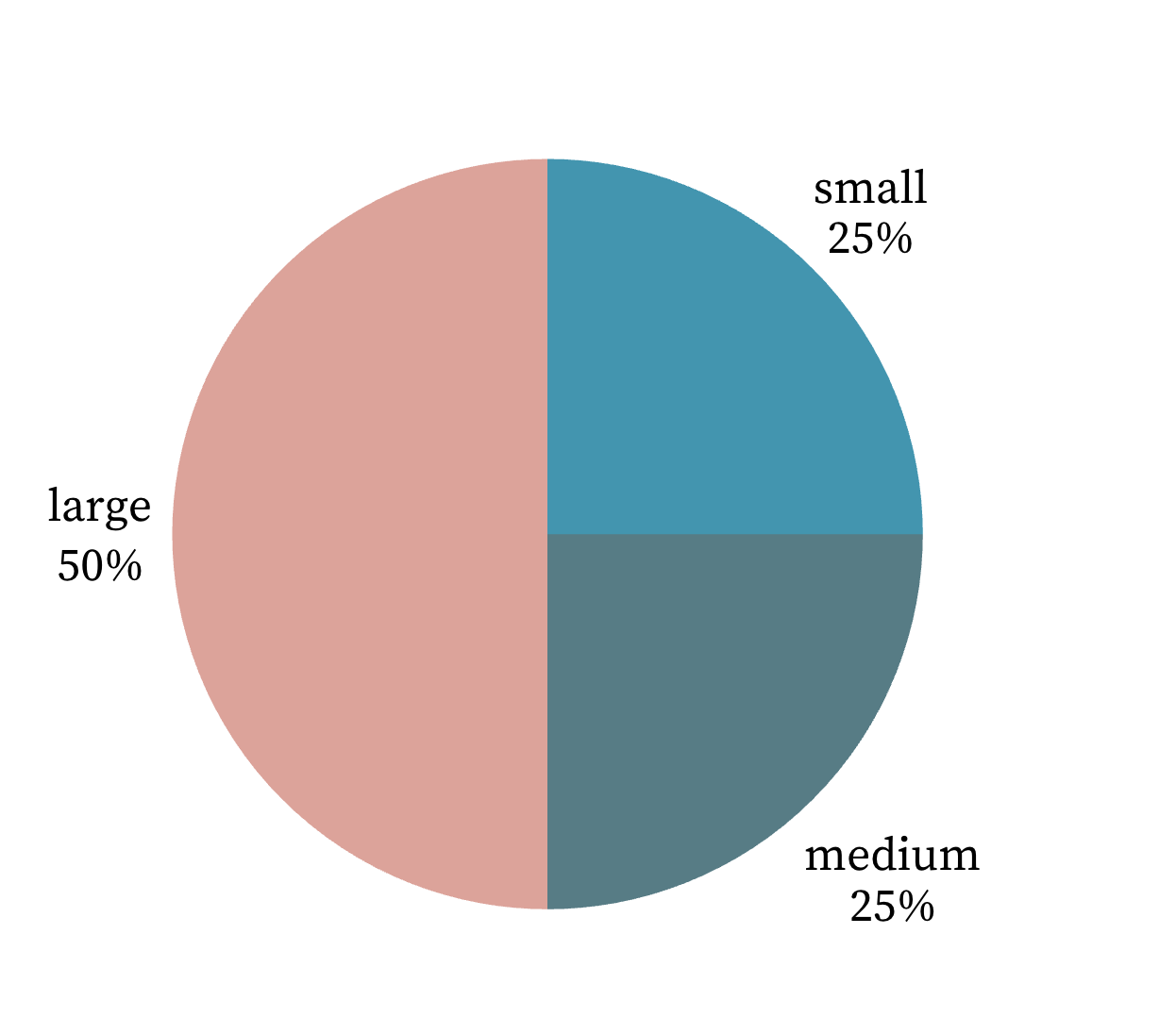}
        \caption{Dataset size distribution.}
        \label{fig:datasetSizeSplit}
    \end{subfigure}
    \hfill
    \begin{subfigure}[t]{0.48\textwidth}
        \centering
        \includegraphics[width=\textwidth]{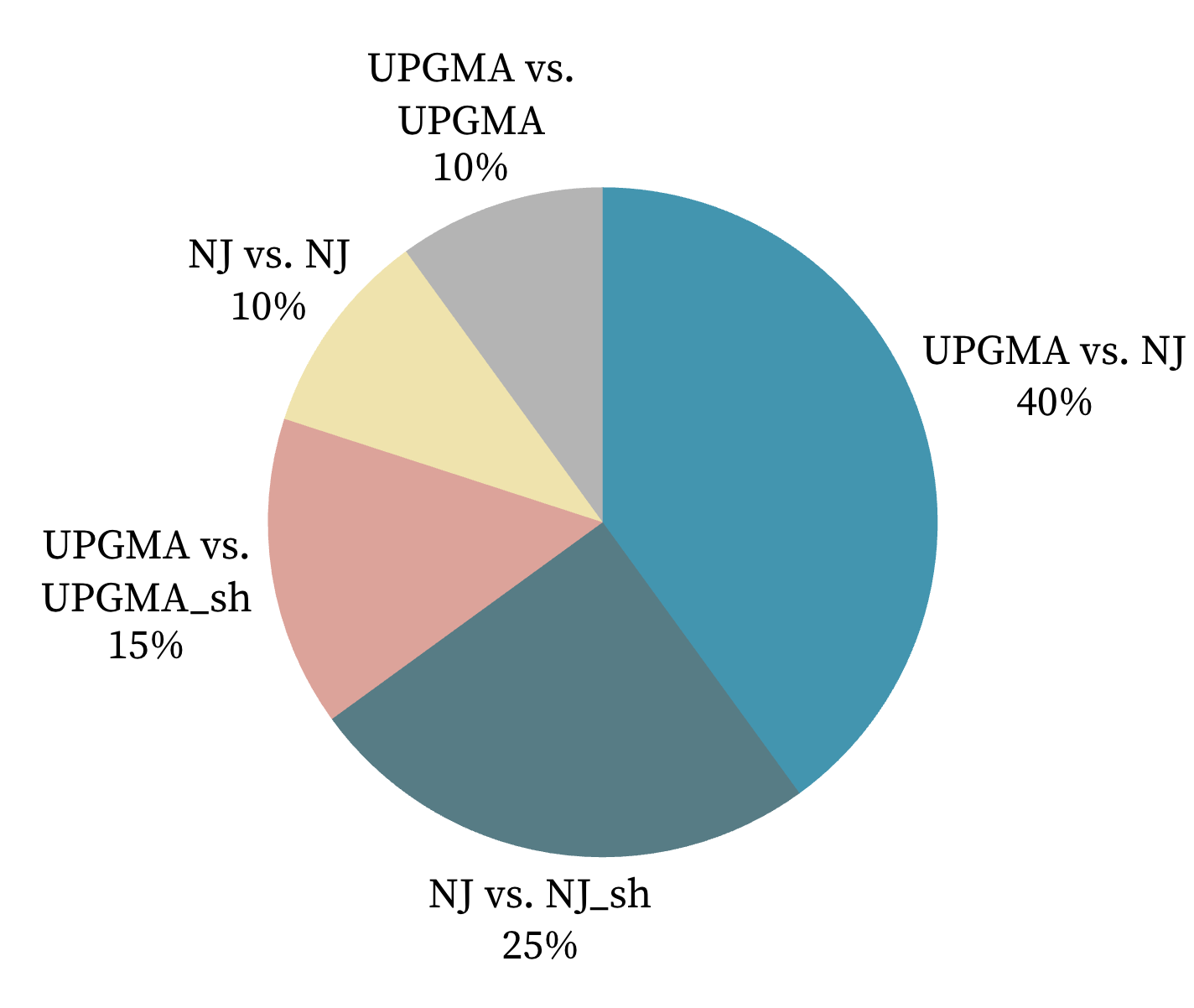}
        \caption{Tree pair distribution for pairing and labelling.}
        \label{fig:treePairDistribution}
    \end{subfigure}
    \caption{Dataset statistics.}
    \label{fig:datasetStatistics}
\end{figure}

We form 388 tree pairs following a fixed niche distribution: UPGMA vs.\
NJ (40\%), NJ vs.\ NJ\textsubscript{sh} (25\%), UPGMA vs.\
UPGMA\textsubscript{sh} (15\%), and two zero-distance control blocks NJ
vs.\ NJ and UPGMA vs.\ UPGMA (10\% each); see
Fig.~\ref{fig:treePairDistribution}. For each pair we extract the SPR
label with \texttt{phangorn::SPR.dist}; the structural trace (the leaves
that move between the two topologies) is obtained from the maximum
agreement forest (MAF) computed exactly with \texttt{rspr}~\cite{whidden2014},
for which the rooted SPR distance equals the number of MAF components
minus one. The master table is split into disjoint train, validation and
test subsets.

Notice that the supervision target is a \emph{heuristic estimate of the unrooted SPR
distance}, the  \texttt{phangorn::SPR.dist}~\cite{deoliveira2008}, and
although it was originally described as a conservative upper-bound
estimate, no error bound was ever formally established, and in practice it
tends to \emph{underestimate} the true SPR distance when that distance is
small relative to tree size, which is precisely our regime.
Second, inspection of the \texttt{phangorn} source shows that
\texttt{SPR.dist} internally \emph{unroots} both inputs (it calls
\texttt{clean\_phylo(\dots, unroot=TRUE)} and canonicalises bipartitions
with \texttt{SHORTwise}) and exposes no \texttt{rooted} option. Passing
rooted trees therefore does not yield a rooted SPR distance. In contrast,
\texttt{rspr}~\cite{whidden2014} computes the \emph{exact rooted} SPR distance. The two tools thus measure related but distinct
quantities. In Section~\ref{sec:eval} we evaluate how the exact rooted SPR
distance compares against the unrooted heuristic estimate from the
\textit{phangorn} package.
\vspace{-0.05cm}
\section{Method}
\label{sec:method}
\begin{figure}[t]
\centering
\includegraphics[width=0.95\textwidth]{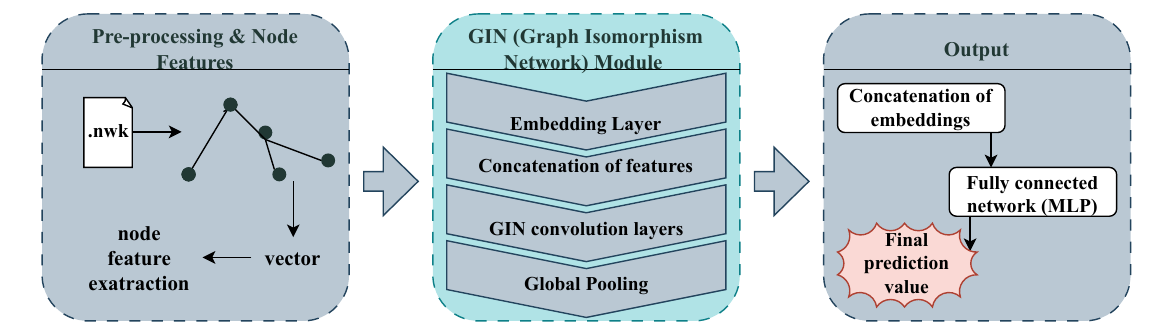}
\caption{High-level architecture. Both trees are encoded by the same
(Siamese) GIN module (node/edge features, GIN convolution with batch
normalisation and ReLU), and the output module pools each tree and
regresses the pair to the predicted SPR distance.}
\label{fig:arch}
\end{figure}

Figure~\ref{fig:arch} gives a high-level overview of the model: a tree
representation module produces node features, a shared GIN module encodes
each tree, and an output module pools and regresses the pair to an SPR
distance.
\subsection{Graph Representation and Node Features}
\begin{figure}[!b]
\centering
\includegraphics[width=0.95\textwidth]{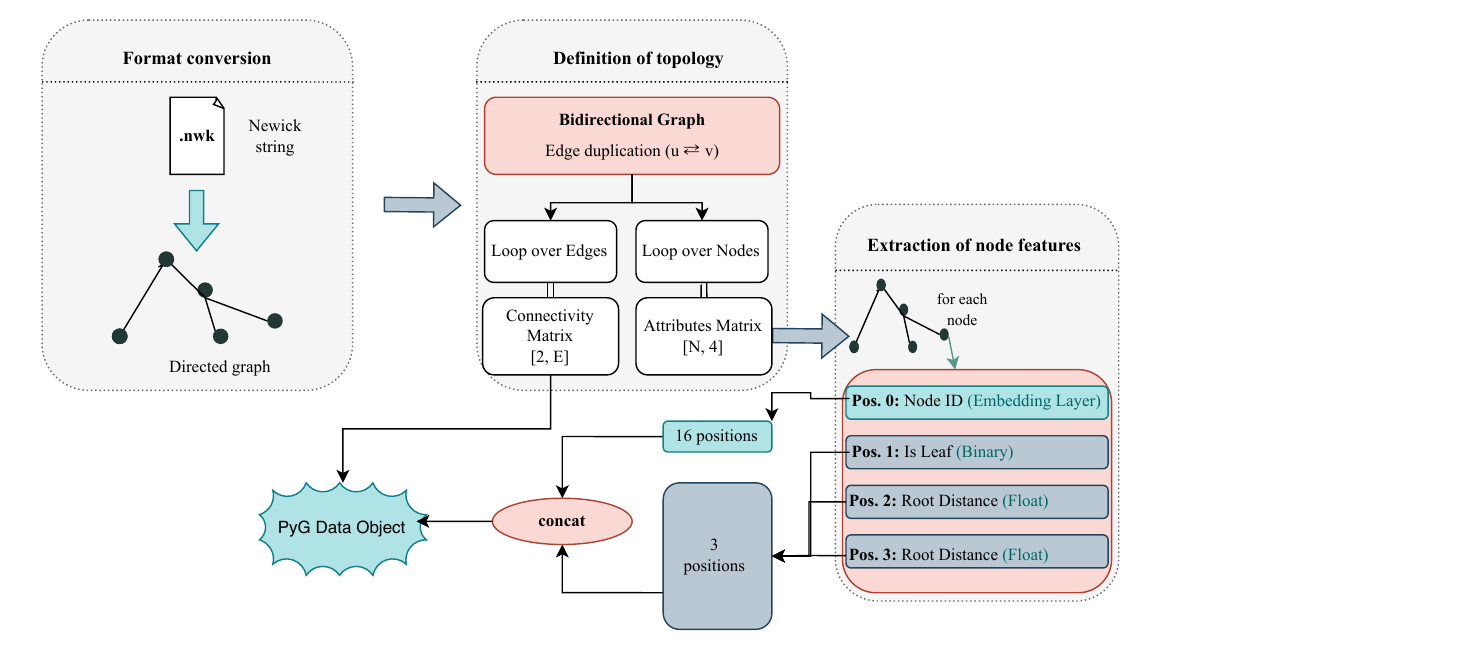}
\caption{Input representation and node encoding. The Newick tree is parsed
into a bidirectional graph; each node yields a feature row
(\emph{degree}, \emph{is-leaf}, \emph{distance to root}, \emph{species id}).
The three structural features (dim 3) are concatenated with a learned
species embedding (dim 16) to form the 19-dimensional node feature.}
\label{fig:encoding}
\end{figure}

Each Newick tree is parsed with Biopython~\cite{cock2009} and converted to
a directed graph whose root is the unique in-degree-zero vertex. Then, it is
encoded as a bidirectional graph in PyTorch Geometric~\cite{fey2019} so
that messages propagate in both directions. In this context, each node carries a
four-dimensional feature vector: node degree, a binary leaf indicator, the
topological distance to the root in number of branches, and a categorical
taxonomic identifier. The first three are used directly, while the
identifier is mapped through a trainable \texttt{Embedding} layer of
dimension 16, with vocabulary size set dynamically to
$\texttt{max\_id}+1$ to handle large isolate counts. Concatenation yields a
19-dimensional input per node. Figure~\ref{fig:encoding} illustrates the
full encoding, from the Newick string to the per-node feature vector.
\subsection{Siamese GIN Regressor}
\begin{figure}[t]
\centering
\includegraphics[width=\textwidth]{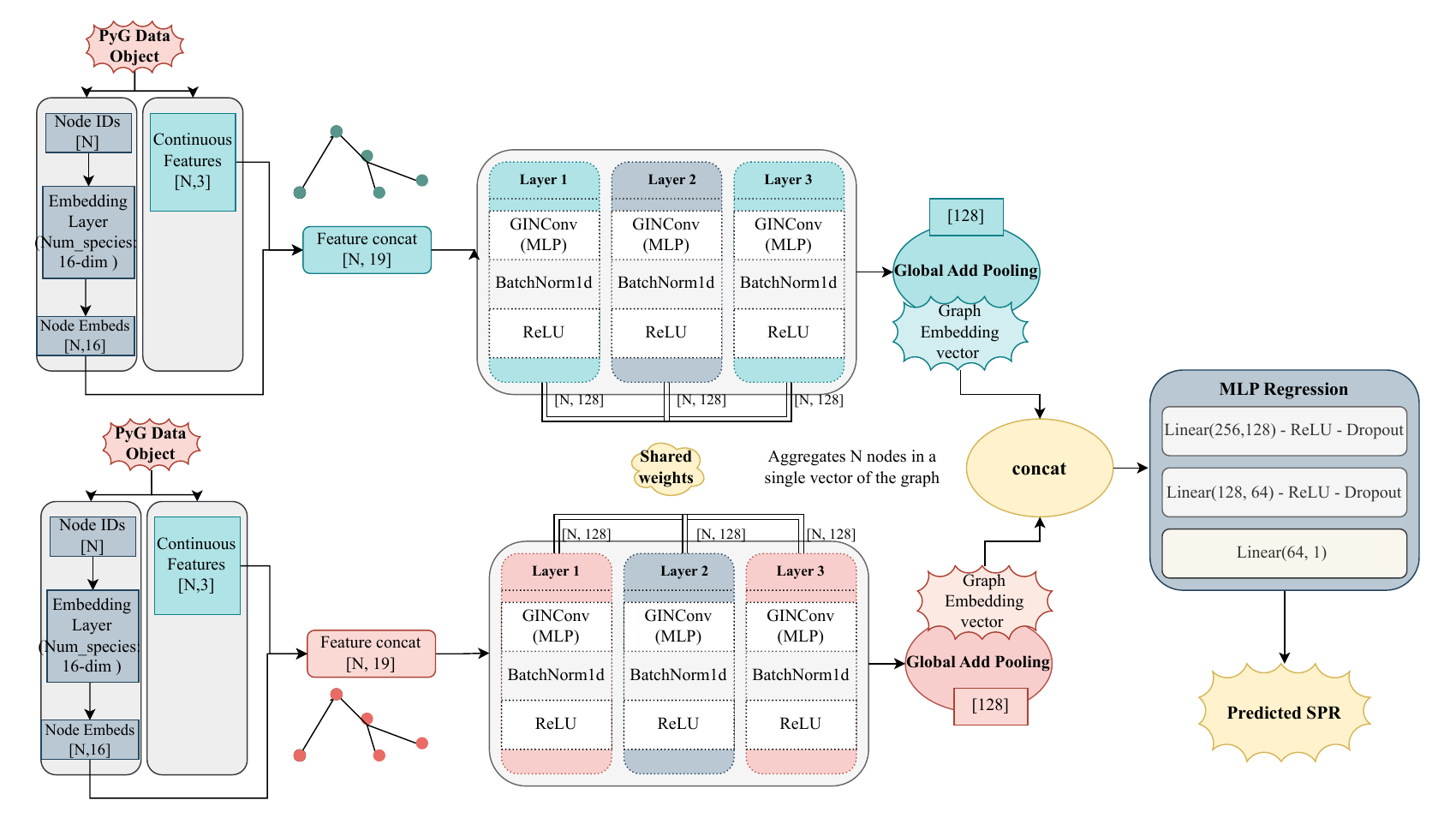}
\caption{Detailed Siamese GIN. Each tree passes through the same encoder
(two GIN layers, each a \texttt{GINConv} MLP \mbox{Linear--ReLU--Linear}
followed by batch normalisation and ReLU); global pooling (\texttt{add}) yields a 128-d embedding per tree. The two embeddings
are concatenated (256-d) and regressed by the MLP head
$256{\to}128{\to}64{\to}1$ to the predicted SPR distance.}
\label{fig:detail}
\end{figure}

The model, detailed in Fig.~\ref{fig:detail}, is a Siamese embedding
network: both trees of a pair pass through the \emph{same} GIN encoder.
Each GIN layer~\cite{xu2019} updates node features as
\begin{equation}
h_i^{(k)} = \mathrm{MLP}^{(k)}\!\left( (1+\epsilon^{(k)})\, h_i^{(k-1)}
+ \sum_{j \in \mathcal{N}(i)} h_j^{(k-1)} \right),
\end{equation}
where $\mathcal{N}(i)$ is the neighbourhood of node $i$. We stack two GIN
layers of hidden dimension 128; each wraps a two-layer MLP,
Linear-ReLU-Linear, followed by batch normalisation~\cite{ioffe2015batch} and
ReLU. A global sum-pooling layer then aggregates node features into a
fixed-size graph embedding
\begin{equation}
v_g = \sum_{i \in V_g} h_i .
\end{equation}
The two embeddings $v_a$ and $v_b$ are concatenated into a 256-dimensional
vector and regressed by an MLP head $256\!\to\!128\!\to\!64\!\to\!1$ with
ReLU activations and dropout $0.3$, producing the predicted SPR distance.
\subsection{Training}
\begin{figure}[t]
\centering
\includegraphics[width=\textwidth]{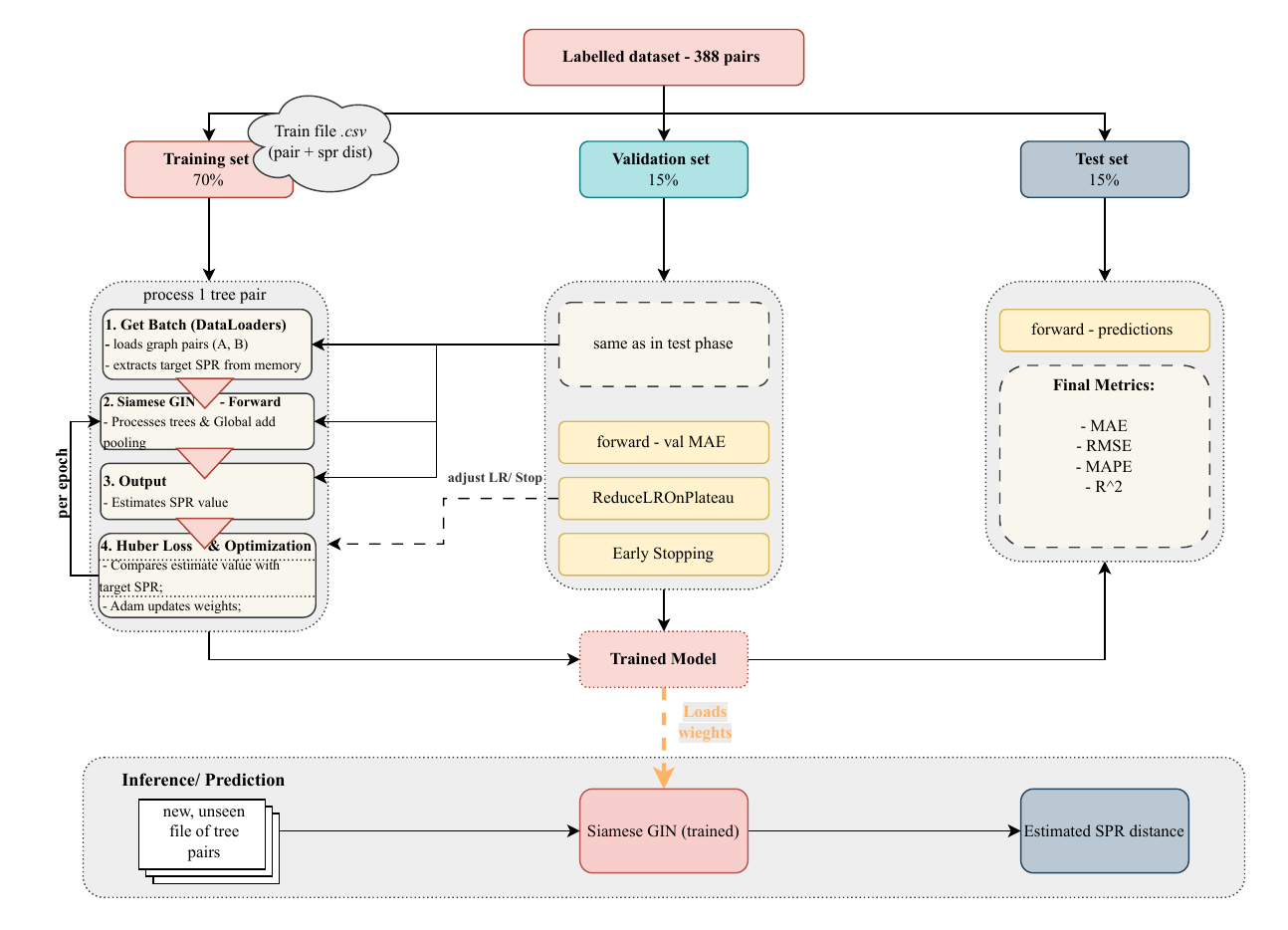}
\caption{Workflow. The labelled pairs are split (70/15/15, seed 42) into
training, validation and test sets. Training back-propagates the Huber
loss with Adam; validation drives learning-rate scheduling and early
stopping and selects the best model; the held-out test set is used once
for the final metrics. At inference, the trained model estimates SPR of a new tree pair in a single forward pass.}
\label{fig:workflow}

\end{figure}

To keep training efficient, each tree is parsed and converted to a PyTorch
Geometric object once and cached in memory, which reduces the per-epoch
cost from $O(\text{epochs}\times N)$ to $O(N)$. The network is trained with
the Adam optimiser~\cite{kingma2015} at a learning rate of $10^{-4}$ and
weight decay of $10^{-4}$, minimising the Huber loss~\cite{huber1964},
which is robust to the right-skewed distribution of SPR values. The
learning rate is halved on plateau after 10 epochs without improvement, and
training stops early after 25 such epochs. We split the labelled pairs
70/15/15 into training, validation and test sets under a fixed seed.
Predicted distances are clamped at zero, since negative distances are
biologically meaningless. Performance is reported as Mean Absolute Error
(MAE), Root Mean Squared Error (RMSE), Mean Absolute Percentage Error
(MAPE) and the coefficient of determination $R^2$. Figure~\ref{fig:workflow}
summarises the full training, validation, testing and inference workflow.

\section{Experimental Evaluation}
\label{sec:eval}
This section evaluates the approach along three axes. We first assess the
scalability of the labelling pipeline; we then quantify how well the
\texttt{phangorn} heuristic used for supervision approximates the exact
rooted SPR distance; and we finally measure the accuracy and
generalization of the trained Siamese GIN across in-distribution,
cross-species, and size-extrapolation regimes.
\subsection{Scalability of the Pipeline}
In stress testing on a virtual machine with 240\,GiB of RAM and Intel
Xeon Silver CPUs, the labelling pipeline scaled to trees of up to
9{,}500 profiles. Exact \texttt{rspr} computation, by contrast, was not feasible for all those trees.
A
trained GNN then predicts a distance in a single forward pass, in a
fraction of the heuristic's time.
\subsection{Validating the Supervision Target}
On small validation trees (20--90 leaves), where exact computation is
tractable, we compared the unrooted \texttt{phangorn} heuristic against the
exact rooted SPR distance. Table~\ref{tab:exact} reports the
\textit{Clostridium} pairs. The heuristic is strongly rank-correlated with
the exact distance (Pearson $0.983$ on \textit{Clostridium}, $0.9935$ on a \textit{Vibrio} subset, and $0.99$ pooled over $n=26$ pairs),
but it systematically \emph{underestimates} the magnitude, returning on
average about $65\%$ for \textit{Clostridium} and $71\%$ for \textit{Vibrio}
of the exact value.
\begin{table}
\centering
\caption{Heuristic (unrooted \texttt{phangorn} estimate) vs.\ exact
rooted SPR (\texttt{rspr}) on the small \textit{Clostridium} validation
pairs (20--90 leaves). \emph{Trace} is the number of relocated leaves,
read from the maximum agreement forest (MAF). Subscript $s$ denotes
shuffled input.}
\label{tab:exact}
\begin{tabular}{c l c c c}
\toprule
\textbf{Size} & \textbf{Pair} & \textbf{Heuristic} & \textbf{Exact} & \textbf{Trace} \\
\midrule
20 & NJ$_s$ vs UPGMA$_s$    & 6  & 9  & 15 \\
30 & UPGMA$_s$ vs UPGMA     & 12 & 20 & 22 \\
40 & UPGMA$_s$ vs UPGMA     & 17 & 28 & 35 \\
45 & UPGMA vs NJ$_s$        & 20 & 33 & 37 \\
50 & UPGMA vs UPGMA$_s$     & 22 & 38 & 43 \\
55 & UPGMA vs NJ            & 5  & 8  & 21 \\
60 & UPGMA vs NJ            & 5  & 9  & 22 \\
65 & UPGMA vs NJ            & 6  & 9  & 26 \\
70 & UPGMA vs NJ            & 7  & 10 & 28 \\
75 & UPGMA vs NJ            & 7  & 11 & 31 \\
80 & UPGMA vs NJ            & 10 & 14 & 33 \\
85 & UPGMA vs NJ            & 14 & 19 & 53 \\
90 & UPGMA vs NJ            & 15 & 20 & 60 \\
\bottomrule
\end{tabular}
\end{table}
This underestimation is expected rather than a heuristic failure, and has
two sources. First, \texttt{phangorn} estimates the \emph{unrooted} SPR
distance, which is generally no larger than the \emph{rooted} distance
preserved by midpoint rooting. Second, the heuristic itself carries no
proven approximation bound and empirically underestimates the true SPR
distance when that distance is small relative to tree size, consistent with
the systematic gap we observe. Both effects push the heuristic below the
exact rooted value. The very high correlation nonetheless shows it is an
excellent \emph{monotonic surrogate}, which is what matters for training a
regressor, even if it is a biased estimator of the rooted magnitude. The
trace (relocated leaves) is far larger than the SPR distance itself, for
example 8 operations but 21 leaves moved at size 55, since a single
prune-and-regraft move can displace an entire subtree.

\subsection{Model Performance}
\begin{figure}[t]
\centering
\includegraphics[width=0.62\textwidth]{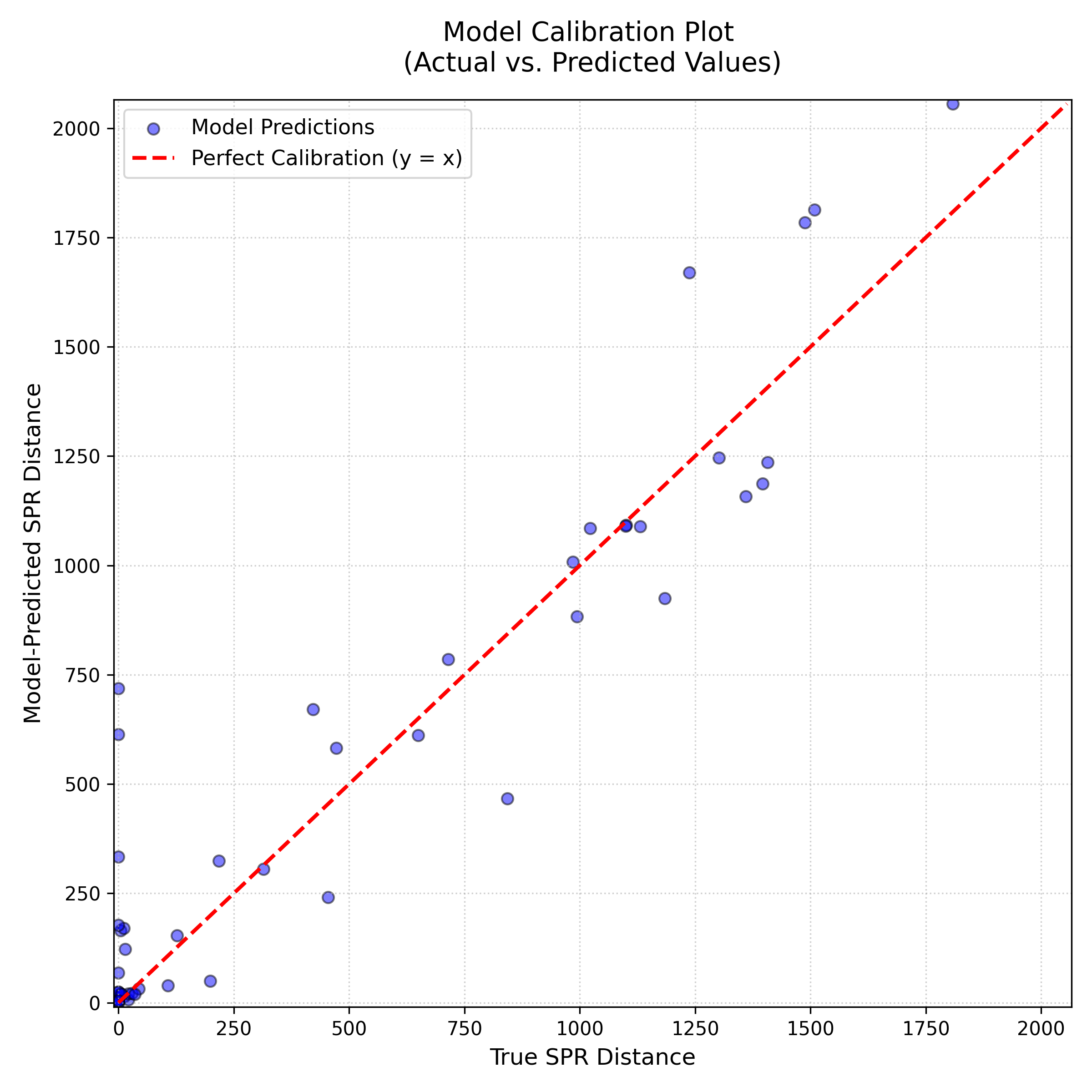}
\caption{Calibration on the in-distribution test set: model-predicted
vs.\ true SPR distance, with the ideal $y=x$ line. Points track the
diagonal but compress towards the mean, with short distances overestimated
and long distances underestimated.}
\label{fig:calibration}
\end{figure}

\begin{figure}[t]
\centering
\includegraphics[width=1.00\textwidth]{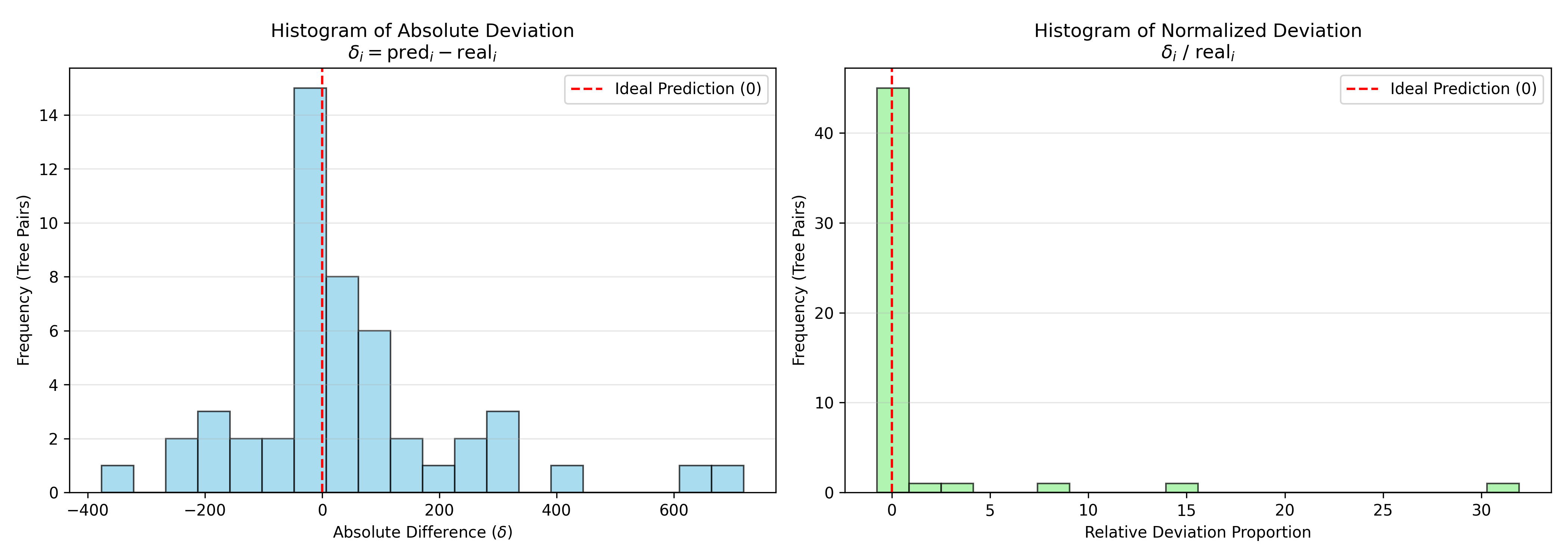}
\caption{Distribution of prediction deviations from the ground-truth SPR
distance. Left: absolute deviation concentrates around zero (ideal
prediction), with a slight right-skew. Right: normalized deviation is
near zero for most instances, indicating that larger absolute deviations
occur mostly at higher baseline distances.}
\label{fig:histogram}
\end{figure}

Under stratified cross-validation the Siamese GIN reaches
$R^2 = 0.905 \pm 0.191$, MAE $= 92.24 \pm 7.02$, RMSE $= 128.14 \pm 11.07$
and MAPE $= 2.16\% \pm 0.37\%$. The standard deviation is driven by
occasional right-skewed outliers in small validation splits rather than by
instability of the fit. Table~\ref{tab:results} reports three single-split
configurations that delimit the model's operating envelope.
\vspace{-0.05cm}
\begin{table}[t]
\centering
\caption{GIN performance across data regimes. ``Baseline'' is the
mean-predictor MAE. In-distribution results are strong; cross-species
transfer is limited and size extrapolation fails.}
\label{tab:results}
\begin{tabular}{lcccc}
\toprule
\textbf{Configuration} & \textbf{MAE} & \textbf{RMSE} & \textbf{$R^2$} & \textbf{Baseline MAE} \\
\midrule
In-distribution (4 species, mixed sizes)  & 127.13 & 202.24 & 0.873  & 485.78 \\
Cross-species (train 2 sp.\ / test 2 sp.) & 208.70 & 325.72 & 0.368  & 406.05 \\
Size extrapolation (small+med.\ $\to$ large) & 375.90 & 643.02 & $-0.14$ & 553.65 \\
\bottomrule
\end{tabular}
\end{table}
Three findings stand out. \emph{(i)} In-distribution, the model explains
$\approx 87\%$ of the variance and beats the mean baseline by roughly
$4\times$ in MAE, so the GNN genuinely learns to compare topologies.
\emph{(ii)} Trained on two species and tested on two unseen ones
(\textit{Clostridium} and \textit{Vibrio} $\to$ \textit{Salmonella} and
\textit{S. pneumoniae}), it attains $R^2 = 0.37$. This indicates that the
learned representation captures some species-agnostic structure, but that
cross-species generalization remains challenging. The drop is hard to
attribute to domain shift alone, since this split also reduces the training
set substantially, so data scarcity is a likely contributing factor.
\emph{(iii)} When trained only on small and medium trees and asked to
predict on large ones, performance falls below the mean baseline
($R^2 = -0.137$): the model does not extrapolate to tree sizes outside its
training range. 
The calibration analysis on the in-distribution test set, shown in
Figure~\ref{fig:calibration}, is consistent with this:
predictions cluster
around the ideal $y=x$ line but compress towards the mean, overestimating
short distances and underestimating long ones.
Figure~\ref{fig:histogram} refines this picture through the deviation histograms.
The absolute-deviation histogram concentrates sharply around zero, so the
model predicts most pairs accurately, with a slight right-skew and a tail
that indicate an occasional tendency to overestimate on complex topologies.
The normalized-deviation histogram is likewise concentrated near zero,
showing that the largest absolute errors fall on pairs with inherently
large SPR distances, so the relative error stays low across most of the
domain.
\vspace{-0.05cm}
\vspace{-0.05cm}
\subsection{Discussion}
Taken together, the experiments show that the binding limitation is neither
the architecture nor, primarily, the absolute size of the labelled set
(388 pairs), but its \emph{coverage} across tree scales. The model
interpolates well and transfers only partially across species, but it
cannot yet extrapolate in size, and it inherits a calibration bias from the
heuristic labels and from sum pooling. These limitations are addressable:
enriching the training distribution with larger trees, exploring size-aware
normalisation, and correcting the label offset are concrete next steps.
Among the pooling strategies we tested, sum pooling gave the best overall
results, but it remains a source of the size-related bias, so alternative
aggregation schemes are worth exploring. 
\vspace{-0.05cm}
\section{Final Remarks}
\label{sec:conclusion}
We presented a reproducible ecosystem for approximating the NP-hard SPR
distance between phylogenetic trees: a pre-processing pipeline, an openly
released dataset of 864 trees and 388 labelled pairs, and a Siamese GIN
regressor with training and inference tools. We validated the supervision
target against exact rooted SPR (Pearson $0.98$--$0.99$) and clarified the
distinction between \texttt{phangorn} and \texttt{rspr}. The model predicts
SPR distances accurately in-distribution ($R^2 \approx 0.87$--$0.90$) and
shows partial transfer across species ($R^2 \approx 0.37$), 
while its current weakness is size extrapolation. Future work will expand
the dataset towards larger topologies, add size-aware pooling and
calibration, and benchmark the learned predictor against exact and
approximate SPR tools.

A further direction is to predict the SPR \emph{trace} rather than only its
magnitude: to identify the specific leaves that must be relocated to
transform one tree into the other, recovering the sequence of
prune-and-regraft operations relating the two topologies. This would let
the model report not only \emph{how far apart} two trees are but
\emph{where} they differ. The trace currently depends on exact MAF
extraction via \texttt{rspr}, whose cost grows prohibitive on large trees,
so trace supervision at scale would itself require a heuristic or learned
approximation of the MAF. Together, these steps target a fast, scalable
approximation suitable for large-scale epidemiological and
comparative-genomic analyses.

%
%
%
%
\begin{credits}
\subsubsection{\ackname}
The authors acknowledge the support by national funds through the
Fundação para a Ciência e a Tecnologia, I.P.\ (FCT) under projects
UID/50021/2025, UID/PRR/50021/2025, UID/PRR/00006/2025, and grant
2023.17447.ICDT.
\subsubsection{\discintname}
The authors have no competing interests to declare that are relevant to
the content of this article.
\end{credits}

%
%
\bibliographystyle{splncs04}
\bibliography{references}
\end{document}